\documentclass[12pt]{article}
\usepackage{times}
\usepackage{geometry}
\usepackage[utf8]{inputenc}
\usepackage{enumitem,amssymb}
\usepackage{ragged2e}
\usepackage{natbib}
\newlist{thematic}{itemize}{8}
\setlist[thematic]{label=$\square$}
\usepackage{pifont}
\usepackage{graphicx}
\newcommand{\cmark}{\ding{51}}%
\newcommand{\done}{\rlap{$\square$}{\raisebox{2pt}{\large\hspace{1pt}\cmark}}%
\hspace{-2.5pt}}

\usepackage{xspace}

\DeclareRobustCommand{\ion}[2]{%
\relax\ifmmode
\ifx\testbx\f@series
{\mathbf{#1\,\mathsc{#2}}}\else
{\mathrm{#1\,\mathsc{#2}}}\fi
\else\textup{#1\,{\mdseries\textsc{#2}}}%
\fi}

\begin{document}
\raggedright
\huge
Astro2020 Science White Paper \linebreak

The fastest components in stellar jets \linebreak
\normalsize

\noindent \textbf{Thematic Areas:} \hspace*{60pt} $\square$ Planetary Systems \hspace*{10pt} $\done$ Star and Planet Formation \hspace*{20pt}\linebreak
$\square$ Formation and Evolution of Compact Objects \hspace*{31pt} $\square$ Cosmology and Fundamental Physics \linebreak
  $\square$  Stars and Stellar Evolution \hspace*{1pt} $\square$ Resolved Stellar Populations and their Environments \hspace*{40pt} \linebreak
  $\square$    Galaxy Evolution   \hspace*{45pt} $\square$             Multi-Messenger Astronomy and Astrophysics \hspace*{65pt} \linebreak
  
\textbf{Principal Author:}

Name:	Hans Moritz G\"unther
 \linebreak						
Institution: Massachusetts Institute of Technology
Kavli Institute for Astrophysics and Space Research 
 \linebreak
Email: hgunther@mit.edu
 \linebreak
Phone:  617-253-8008
 \linebreak
 
\textbf{Co-authors:} (names and institutions)\\
Catherine Espaillat (Boston University),
Kevin France (University of Colorado),
Zhi-Yun Li (University of Virginia), Christopher M. Johns--Krull (Rice University), Catherine Dougados (IPAG, France),
P. Christian Schneider (Hamburg Observatory), Will Fischer (STScI), Scott
J. Wolk (SAO), Tracy L. Beck (STScI), Nuria Calvet (University of Michigan),
Manuel G\"udel (University of Vienna)
  \linebreak

\textbf{Abstract:}
Young stars accrete mass from a circumstellar disk, but at the same time disk
and star eject outflows and jets. These outflows have an onion-like structure
where the innermost and fastest layers are surrounded by increasingly lower
velocity components. The outer layers are probably photo-evaporative and magnetocentrifugally launched disk winds, but the nature of the inner winds is still uncertain. Since the fastest components carry only a small fraction of the mass, they are best observed at high-energies (X-ray and UV) as the  slower, more massive components do not reach plasma temperatures sufficient for relevant X-ray or UV emission. Outflows are the most likely way in which a star or its disk can shed angular momentum and allow accretion to proceed; thus we cannot understand the accretion and the rotation rate of young stars if we cannot solve the origin of the inner jet components. Stellar jets share characteristics with their counterparts in more massive astrophysical objects, such as stellar mass black holes and AGN, with the added benefit that young stars are found at much closer distances and thus scales not accessible in other types of objects can be resolved.

To understand the origin and impact of the inner jets, sub-arcsecond imaging and spectroscopy in the UV and X-rays is required, together with theory and modelling to interpret existing and future observations. 

\setlength{\parindent}{2em}

\pagebreak
\section{Jets from young stars}
Stars form when giant molecular clouds fragment and contract into
proto-stars. Mass accretion onto those stellar cores proceeds via an accretion
disk, while the surrounding envelope eventually disperses. In this stage, the
stars become visible at all wavelengths; the low-mass population is called
classical T~Tauri stars (CTTS), the A and B star progenitors are Herbig Ae/Be
stars (HAeBe). Jets and outflows are a natural consequence of disk
accretion. They are not only common phenomena in star formation, but have been
detected in most classes of accreting objects ranging from AGN through CV
systems down to brown dwarfs. Outflows come in different shapes: There are slow
wide-angle winds, often seen in molecular lines (e.g. H$_2$, or CO), faster winds,
showing up in forbidden optical lines such as [O~{\sc i}] 6300~\AA{}, and highly collimated jets, which often reach velocities up to 400~km~s$^{-1}$ \citep{1998AJ....115.1554E}. 

Different theoretical models of stellar winds \citep{1988ApJ...332L..41K,2005ApJ...632L.135M}, X-winds \citep{1994ApJ...429..781S} and disk winds \citep{1982MNRAS.199..883B,2005ApJ...630..945A} have been proposed. Ultimately the jet launching must be powered from the gravitational energy released in the accretion process. This is supported by the observation that the outflow rate is roughly one tenth of the accretion rate \citep{1990ApJ...354..687C,2008ApJ...689.1112C}, but it is unclear how the energy is converted. 

\begin{figure}[htb]
\centering
\includegraphics[width=\textwidth]{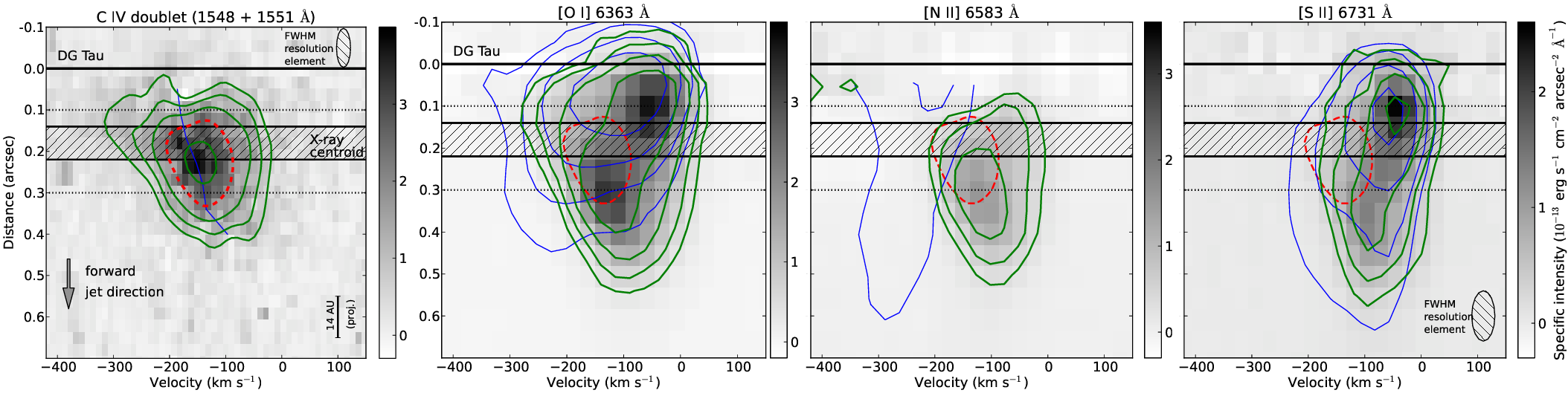}
\caption{Position-velocity diagrams (PVD) from long-slit observations with HST/STIS of the CTTS DG~Tau. The figure nicely shows how data from the X-ray to the optical is required to study how the different jet components are related to each other. The red dashed contour indicates the C~{\sc iv} emission and the horizontal dotted lines give the peak location of the two optical knots. The shaded area indicates the centroid of the inner X-ray jet, the blue line visualizes the velocity of the C~{\sc iv} emission, and the blue contours pertain to the central jet emission in an earlier epoch. From: \protect{\citet{2013A&A...550L...1S}}}
\label{fig:CIV}
\end{figure}

\begin{figure}[htb]
\centering
\includegraphics[height=6cm]{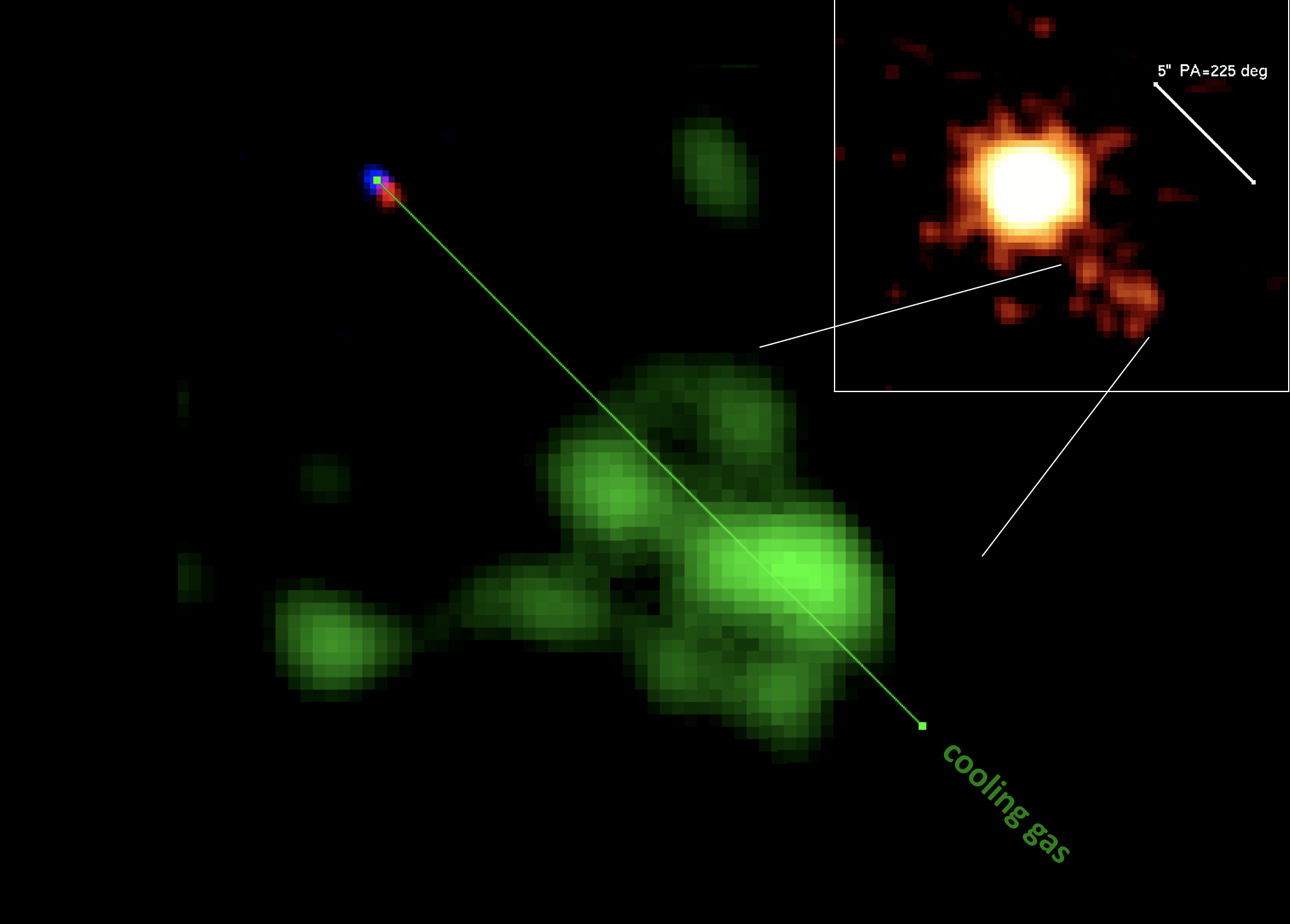}
\includegraphics[height=6cm]{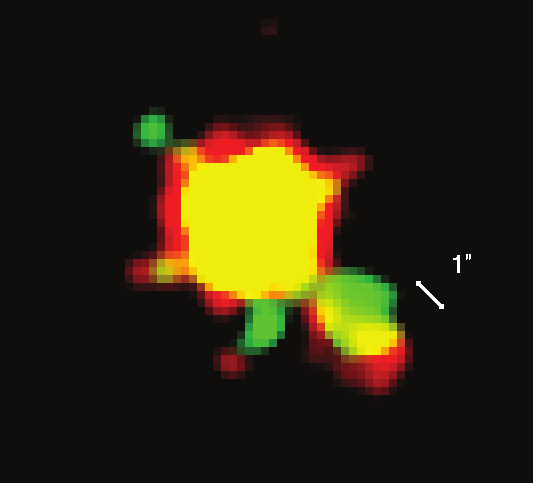}
\caption{\emph{left:} The jet from the CTTS DG Tau with \emph{Chandra}. The
inset shows the conventional image, where the jet at a few hundred au is
resolved, but the inner component is not seen. For the main figure, a
deconvolution algorithm was applied, showing hard emission from the star
(blue) and the inner jet in red (note the shift by about 0.15”). This looks
good, but the result is biased because the algorithm is insensitive between
the red part and the green outer jet. To resolve the components seen in the
UV and optical in Fig.~\ref{fig:CIV}, the native image resolution needs to be
better than \emph{Chandra}. \emph{Right}: Two-color image of the CTTS DG Tau and its jet (towards the bottom right) and counterjet (barely visible towards the top left), showing superposition of smoothed X-ray
images for Winter 2005/06 (green) and January 2010 (red), indicating jet
longitudinal motion. From: \protect{\citet{2011ASPC..448..617G}}}
\label{fig:Xray}
\end{figure}

We currently believe that the jets are layered like an onion where slow and
cool outflow components surround successive layers of faster and more
collimated jet components. The outermost layers are disk winds launched tens of
AU from the central star, possibly through photo-evaporation from the irradiated disk surface, but it is unclear where the innermost components come
from. The mass flux in these innermost components is no more than $10^{-3}$ \citep{2009A&A...493..579G} of
the total jet mass flux, but they are fast enough to generate X-ray emission.
While the launching region of this inner jet is (and will remain in the 2020s) spatially unresolved, we can see an inner \citep[10-40 AU, Fig.~\ref{fig:CIV},][]{2008A&A...488L..13S} emission region probably associated with jet collimation, as well as an outer \citep[hundreds of AU, Fig.~\ref{fig:Xray},][]{2011ASPC..448..617G} region powered by shocks between different jet components or between the jet and the interstellar medium. 

\section{Questions for the 2020s}
In the last two decades, \emph{Chandra} and \emph{HST/STIS} have slowly chipped away at observing one stellar jet at a time, leading to a sample of a handful of objects with a few cases where the timeline is sufficient to see knots in the jet move \citep[e.g.][]{2011A&A...530A.123S}. However, it has proven hard to draw general conclusions about jet properties and jet launching from this limited set of data, and thus the big questions of jet launching are still open for the next decade:

\subsection{Do jets rotate?}
Disk accretion cannot proceed without the redistribution of angular momentum, either within the disk or within the system. Only recently simulations became mature enough to demonstrate that mechanisms that have long been favored for angular momentum removal cannot operate in important regions of the disks, e.g.\ the magneto-rotational instability (MRI) shuts down in the ``dead zone''. This  leaves jets as the best known alternative to remove angular momentum from these regions, but an observational test of this idea is still outstanding since results for jet rotation are so far inconclusive due to limited exposure times and sample sizes \citep{2007ApJ...663..350C}. Magneto-centrifugally launched jets originate in the inner region around the central object, within a few AU for typical stars. Since post-shock temperatures scale with $v^2_\mathrm{shock}$, the inner most region crucial for the interaction of the accreting object and the disk, requires observations in the UV range where the dominant cooling agents are situated.
As the extinction increases towards shorter wavelength, the contrast between the star and the jet, which is less extincted than the star itself, is much
more favorable in the UV.

\subsection{How do jets launch and collimate?}
Jet width and velocity increase with distance to the star. The exact profile depends on the magnetic field and the conditions in the launching region; to turn this around, if we can measure velocity profiles in the jet (Fig.~\ref{fig:CIV}), we can infer the conditions around the launching radius which will remain inaccessible to direct observations even for the largest telescopes becoming available in the 2020s.

\subsection{Do all jets have a fast inner component?}
To study the interaction between disk, accretion streams and the star, observations of the fastest jet components are most valuable. If we find velocities of 500~km/s and assume
reasonable values for the toroidal jet velocity, the launching region must be
within $10\;R_*$ according to \citet{2003ApJ...590L.107A}. This approximately
corresponds to the inner edge of the disk and thus would rule out a disk
wind. However, with current instrumentation we are limited to study the X-ray
and UV emission of only the brightest sources. We do not know if the mechanism
that accelerates the innermost jet operates in every source (and we just do not
see it because it is faint) or not. With an increased sample size, we could search for correlations of the fast jet components with accretion rate, stellar rotation rate, or stellar magnetic field strength and thus identify the launching mechanism.

\subsection{How does jet launching work in objects with weak magnetic fields?}
All models for jet launching rely on large-scale, ordered magnetic fields \citep{2009ASSP...13...99F}, but jets are observed not only from solar-mass CTTS, which have strong magnetic fields in the kG range. HAeBes also launch jets, although only primordial magnetic fields are expected there, and often only weak fields, if any, can be detected \citep{2007A&A...463.1039H,2007MNRAS.376.1145W}. For example, the formal limit on the magnetic field of HD~163296 is $-25\pm 27$~G \citep{2007A&A...463.1039H}.

\section{Required capabilities and international context}
To probe all the layers of the ``onion'', we need observations in a large
wavelength range from the radio up to X-rays. For all observations, the three
key factors are (1) spatial resolution to actually resolve the jet and features
like shock surfaces within, (2) sufficient signal-to-noise to identify weak, extended
emission, and (3) spectral resolution to measure the kinematics of the emission
lines. The better an observatory does on these metrics, the finer detail can be
resolved and the more jets are accessible to study. Several star forming
regions can be found in about 150~pc distance. For those, we need resolutions
on the level of about 0.1-0.5 arcseconds to resolve the jet perpendicular to
its axis and to identify shock fronts and knots; better spatial resolution
could reveal sub-structure in the jet components, if it exists. Imaging is
good, but it does not tell us where the gas moves, if the jet rotates, and in
which direction a shock expands. To this end, we need to measure the centroid and
width of an emission line, or, even better, resolve it into several kinematic
components. Turbulence probably broadens all emission lines in jets to some
degree, so there is little additional benefit of resolving lines into more than
about ten velocity bins. For the slow moving jet layers seen in the radio or the IR, this requires resolution on the level of a few km/s, for the faster flows seen in the UV and X-rays, a resolution of several 10s of km/s is sufficient.

\subsection{Observational capabilities}
For all wavelengths accessible from the ground, great progress has been made in the last decade and projects well into the planning stages or even in construction, like the US- or ESO led giant telescopes, will provide instruments that improve significantly over existing instrumentation. Integral field units (IFUs) coupled with adaptive optics are the workhorses for jet observations in the IR and optical, and the next generation of these instruments as planned for the extremely large telescopes match the requirements spelled out above. On the other hand, in the UV and X-rays, the situation looks dire. 

No X-ray instrument ever flown had the capability to kinematically resolve jet emission lines within reasonable integration times. Worse, jet observations in X-rays are almost impossible today, since Chandra is the only instrument capable of resolving the jets from their central stars and increasing contamination on the detectors has reduced Chandra's sensitivity to soft X-rays by several magnitudes compared to launch. Upcoming X-ray observatories like Athena or e-ROSITA prioritize collecting area over spatial resolution and will not be able to resolve stellar jets from their central stars. \textbf{New X-ray instrumentation with order of magnitude improvements over exiting instruments is needed to spatially and spectrally resolve stellar jets.}

In the UV, HST/STIS is currently the only instrument available for observations of stellar jets; some examples are presented above where STIS is used to resolve kinematic structure in the jet using long-slit observations. However, without an IFU, we are limited to a single slice of the jet per observation. It is possible to step the long-slit on the sky, but this is expensive in terms of observing time and sample studies are challenging. In the UV, we have shown how much physical information we can extract from the spectra, but crucial questions like jet rotation and comparative studies of a major sample of jets are just outside the capabilities of the current instrumentation. \textbf{New UV instrumentation with a modest improvement in spectral and spatial resolution is useful, but most important is an increase in sample size and observing efficiency (e.g. by using an IFU) such that a sample of jets can be monitored in time. }

\subsection{Theory and Modelling}
\begin{figure}[htb]
\centering
\includegraphics[width=0.5\textwidth]{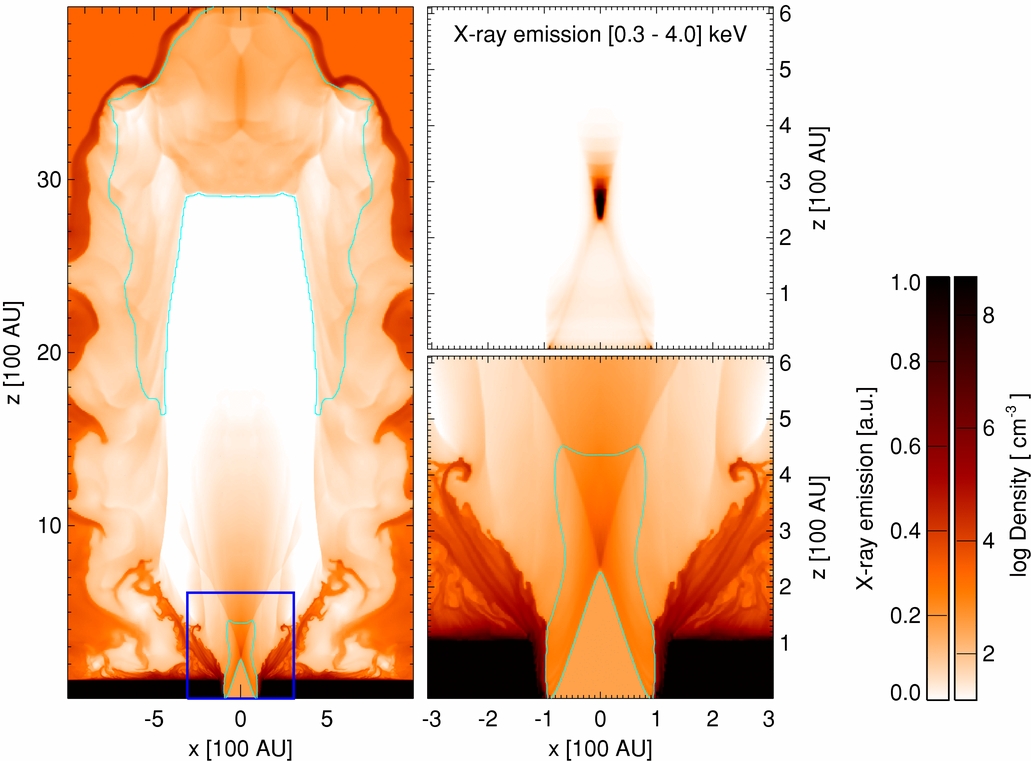}
\caption{Numerical simulation of a shock in a protostellar jet. Shown is a density map (left panel), an enlargement of the base of the computational domain, and the X-ray map synthesized from the model (upper panel on the right).  From: \protect{\citet{2011ApJ...737...54B}}}
\label{fig:sims}
\end{figure}

While the authors of this white paper are mostly observers, we realize the
crucial role that theory and modelling play in understanding stellar
jets. Explaining jet launching and collimation has been an issue in MHD models
for quite some time and several groups are working on that; similarly jets
propagating into the interstellar medium have been simulated in different
contexts, but so far those models use pretty simple jet cross-sections,
e.g.\ just a top-hat profile. Only one model for the X-ray generation in
stellar jets, the diamond shock (see Fig.~\ref{fig:sims}), has been
investigated in detail in numerical simulations, a second model has been looked at at least semi-analytically. While the diamond shock does explain the main observed properties of the X-ray emission at the low-resolution we have so far, it may not be the only model to do so. It will be a challenge for the community in the coming decade to systematically confirm or rule-out jet models through detailed numerical simulations and comparison of the simulations with existing and future observations. \textbf{Thus, theoretical work needs supported together with observations by providing infrastructure, financial support for Post-Docs and staff, and mechanisms that increase the interaction between observers and theorists.}

\pagebreak
\bibliographystyle{aasjournal}
\bibliography{bibl}

\begin{thebibliography}{}
\expandafter\ifx\csname natexlab\endcsname\relax\def\natexlab#1{#1}\fi
\providecommand{\url}[1]{\href{#1}{#1}}
\providecommand{\dodoi}[1]{doi:~\href{http://doi.org/#1}{\nolinkurl{#1}}}
\providecommand{\doeprint}[1]{\href{http://ascl.net/#1}{\nolinkurl{http://ascl.net/#1}}}
\providecommand{\doarXiv}[1]{\href{https://arxiv.org/abs/#1}{\nolinkurl{https://arxiv.org/abs/#1}}}

\bibitem[{{Anderson} {et~al.}(2003){Anderson}, {Li}, {Krasnopolsky}, \&
  {Blandford}}]{2003ApJ...590L.107A}
{Anderson}, J.~M., {Li}, Z.-Y., {Krasnopolsky}, R., \& {Blandford}, R.~D. 2003,
  \apjl, 590, L107, \dodoi{10.1086/376824}

\bibitem[{{Anderson} {et~al.}(2005){Anderson}, {Li}, {Krasnopolsky}, \&
  {Blandford}}]{2005ApJ...630..945A}
---. 2005, \apj, 630, 945, \dodoi{10.1086/432040}

\bibitem[{{Blandford} \& {Payne}(1982)}]{1982MNRAS.199..883B}
{Blandford}, R.~D., \& {Payne}, D.~G. 1982, \mnras, 199, 883,
  \dodoi{10.1093/mnras/199.4.883}

\bibitem[{{Bonito} {et~al.}(2011){Bonito}, {Orlando}, {Miceli}, {Peres},
  {Micela}, \& {Favata}}]{2011ApJ...737...54B}
{Bonito}, R., {Orlando}, S., {Miceli}, M., {et~al.} 2011, \apj, 737, 54,
  \dodoi{10.1088/0004-637X/737/2/54}

\bibitem[{{Cabrit} {et~al.}(1990){Cabrit}, {Edwards}, {Strom}, \&
  {Strom}}]{1990ApJ...354..687C}
{Cabrit}, S., {Edwards}, S., {Strom}, S.~E., \& {Strom}, K.~M. 1990, \apj, 354,
  687, \dodoi{10.1086/168725}

\bibitem[{{Coffey} {et~al.}(2008){Coffey}, {Bacciotti}, \&
  {Podio}}]{2008ApJ...689.1112C}
{Coffey}, D., {Bacciotti}, F., \& {Podio}, L. 2008, \apj, 689, 1112,
  \dodoi{10.1086/592343}

\bibitem[{{Coffey} {et~al.}(2007){Coffey}, {Bacciotti}, {Ray}, {Eisl{\"o}ffel},
  \& {Woitas}}]{2007ApJ...663..350C}
{Coffey}, D., {Bacciotti}, F., {Ray}, T.~P., {Eisl{\"o}ffel}, J., \& {Woitas},
  J. 2007, \apj, 663, 350, \dodoi{10.1086/518100}

\bibitem[{{Eisl{\"o}ffel} \& {Mundt}(1998)}]{1998AJ....115.1554E}
{Eisl{\"o}ffel}, J., \& {Mundt}, R. 1998, \aj, 115, 1554,
  \dodoi{10.1086/300282}

\bibitem[{{Ferreira}(2009)}]{2009ASSP...13...99F}
{Ferreira}, J. 2009, Astrophysics and Space Science Proceedings, 13, 99,
  \dodoi{10.1007/978-3-642-00576-3_12}

\bibitem[{{G{\"u}del} {et~al.}(2011){G{\"u}del}, {Audard}, {Bacciotti}, {Bary},
  {Briggs}, {Cabrit}, {Carmona}, {Codella}, {Dougados}, {Eisl{\"o}ffel},
  {Gueth}, {G{\"u}nther}, {Herczeg}, {Kundurthy}, {Matt}, {Mutel}, {Ray},
  {Schmitt}, {Schneider}, {Skinner}, \& {van Boekel}}]{2011ASPC..448..617G}
{G{\"u}del}, M., {Audard}, M., {Bacciotti}, F., {et~al.} 2011, in Astronomical
  Society of the Pacific Conference Series, Vol. 448, 16th Cambridge Workshop
  on Cool Stars, Stellar Systems, and the Sun, ed. C.~{Johns-Krull}, M.~K.
  {Browning}, \& A.~A. {West}, 617

\bibitem[{{G{\"u}nther} {et~al.}(2009){G{\"u}nther}, {Matt}, \&
  {Li}}]{2009A&A...493..579G}
{G{\"u}nther}, H.~M., {Matt}, S.~P., \& {Li}, Z.-Y. 2009, \aap, 493, 579,
  \dodoi{10.1051/0004-6361:200810886}

\bibitem[{{Hubrig} {et~al.}(2007){Hubrig}, {Pogodin}, {Yudin}, {Sch{\"o}ller},
  \& {Schnerr}}]{2007A&A...463.1039H}
{Hubrig}, S., {Pogodin}, M.~A., {Yudin}, R.~V., {Sch{\"o}ller}, M., \&
  {Schnerr}, R.~S. 2007, \aap, 463, 1039, \dodoi{10.1051/0004-6361:20066090}

\bibitem[{{Kwan} \& {Tademaru}(1988)}]{1988ApJ...332L..41K}
{Kwan}, J., \& {Tademaru}, E. 1988, \apjl, 332, L41, \dodoi{10.1086/185262}

\bibitem[{{Matt} \& {Pudritz}(2005)}]{2005ApJ...632L.135M}
{Matt}, S., \& {Pudritz}, R.~E. 2005, \apjl, 632, L135, \dodoi{10.1086/498066}

\bibitem[{{Schneider} {et~al.}(2013){Schneider}, {Eisl{\"o}ffel}, {G{\"u}del},
  {G{\"u}nther}, {Herczeg}, {Robrade}, \& {Schmitt}}]{2013A&A...550L...1S}
{Schneider}, P.~C., {Eisl{\"o}ffel}, J., {G{\"u}del}, M., {et~al.} 2013, \aap,
  550, L1, \dodoi{10.1051/0004-6361/201118592}

\bibitem[{{Schneider} {et~al.}(2011){Schneider}, {G{\"u}nther}, \&
  {Schmitt}}]{2011A&A...530A.123S}
{Schneider}, P.~C., {G{\"u}nther}, H.~M., \& {Schmitt}, J.~H.~M.~M. 2011, \aap,
  530, A123, \dodoi{10.1051/0004-6361/201016305}

\bibitem[{{Schneider} \& {Schmitt}(2008)}]{2008A&A...488L..13S}
{Schneider}, P.~C., \& {Schmitt}, J.~H.~M.~M. 2008, \aap, 488, L13,
  \dodoi{10.1051/0004-6361:200810261}

\bibitem[{{Shu} {et~al.}(1994){Shu}, {Najita}, {Ostriker}, {Wilkin}, {Ruden},
  \& {Lizano}}]{1994ApJ...429..781S}
{Shu}, F., {Najita}, J., {Ostriker}, E., {et~al.} 1994, \apj, 429, 781,
  \dodoi{10.1086/174363}

\bibitem[{{Wade} {et~al.}(2007){Wade}, {Bagnulo}, {Drouin}, {Landstreet}, \&
  {Monin}}]{2007MNRAS.376.1145W}
{Wade}, G.~A., {Bagnulo}, S., {Drouin}, D., {Landstreet}, J.~D., \& {Monin}, D.
  2007, \mnras, 376, 1145, \dodoi{10.1111/j.1365-2966.2007.11495.x}

\end{thebibliography}

\end{document}